**ORIGINAL ARTICLE**

# Data augmentation for generating synthetic electrogastrogram time series

Nadica Miljković[1,2] · Nikola Milenić[1] · Nenad B. Popović[1] · Jaka Sodnik[2]

**Abstract**
To address an emerging need for large number of diverse datasets for rigor evaluation of signal processing techniques, we developed and evaluated a new method for generating synthetic electrogastrogram time series. We used electrogastrography (EGG) data from an open database to set model parameters and statistical tests to evaluate synthesized data. Additionally, we illustrated method customization for generating artificial EGG time series alterations caused by the simulator sickness. Proposed data augmentation method generates synthetic EGG data with specified duration, sampling frequency, recording state (postprandial or fasting state), overall noise and breathing artifact injection, and pauses in the gastric rhythm (arrhythmia occurrence) with statistically significant difference between postprandial and fasting states in > 70% cases while not accounting for individual differences. Features obtained from the synthetic EGG signal resembling simulator sickness occurrence displayed expected trends. The code for generation of synthetic EGG time series is not only freely available and can be further customized to assess signal processing algorithms but also may be used to increase data diversity for training artificial intelligence (AI) algorithms. The proposed approach is customized for EGG data synthesis but can be easily utilized for other biosignals with similar nature such as electroencephalogram.

**Abbreviations**

| | |
|---|---|
| AI | Artificial intelligence |
| a.u. | Arbitrary unit |
| BMI | Body mass index |
| C | Copyright |
| CF | Crest factor of PSD |
| cpm | Cycles per minute |
| DL | Deep learning |
| DF | Dominant frequency |
| ECA | Electrical control activity |
| ERA | Electrical response activity |
| ECG | Electrocardiogram |
| EGG | Electrogastrography |
| EMG | Electromyography |
| IFFT | Inverse fast Fourier transform |
| FSD | Percentage of the high power spectrum density |
| GPL | General public license |
| GAN | Generative adversarial network |
| ML | Machine learning |
| MF | Median frequency of PSD |
| RMS | Root mean square |
| RSA | Running spectrum analysis |
| PSD | Power spectral density |
| SD | Standard deviation |
| syEGG | Synthetic electrogastrogram time series |
| XAI | Explainable artificial intelligence |

## 1 Introduction

Recording large amount of data, especially in patients, is time consuming and can be challenging due to the privacy issues, as well as on patients' tiredness and impairment. On the other hand, rigor evaluation of signal processing algorithms is in a need for diverse datasets. Mimicking and reproducing real medical data is inevitable for proper quantitative evaluation of image and signal processing workflows, especially in the presence of noise. Also, prevailing machine learning (ML) models are vastly depended on datasets and

✉ Nadica Miljković
nadica.miljkovic@etf.bg.ac.rs

1 University of Belgrade–School of Electrical Engineering, Bulevar Kralja Aleksandra 73, 11000 Belgrade, Serbia

2 Faculty of Electrical Engineering, University of Ljubljana, Tržaška Cesta 25, 1000 Ljubljana, Slovenia

extensive amount of data is required for proper training of ML models. Therefore, diverse and annotated datasets are paramount for addressing ML overfitting due to the class imbalance that is typically pronounced in clinical studies, especially in those involving rare medical conditions. The related data shortage in medicine and healthcare is commonly solved by generating synthetic and realistic datasets by data augmentation. Mimicking and reproducing real medical data could be also used for proper quantitative evaluation of signal processing workflows, especially in the presence of noise [1–14].

Biomedical signal synthetic generators utilize physics-based, statistical, or state-of-the-art deep learning (DL) approaches [1, 15]. For example, DL-based generative model termed generative adversarial network (GAN) has been proposed to create realistic synthetic biomedical signals, especially electrocardiogram (ECG) [1, 5, 16]. Attractive GAN approach proved useful for simulation of artificial data by learning the distribution from the real-world training dataset [1, 8]. However, GANs come at the price of being unstable during training; they do not have proper evaluation metrics [2] and cannot be used to directly control dataset characteristics and labels that comply with the reference medical standards and prior data knowledge [8, 12].

Unlike widely available biomedical signals such as ECG that are customarily hosted on PhysioNet[1] web-based databases [1, 17, 18] or other repositories, open access datasets with electrogastrography (EGG)[2] time series are scarce. Currently available databases[3] comprise the following: (1) three-channel EGG signal recordings in 20 healthy individuals acquired for 20 min of both fasting and postprandial [21, 22], (2) raw EGG signals after drinking saline at different temperatures from 60 subjects [23], (3) features derived from EGG data for the study of phase locking value between resting state network and EGG signals [24], and (4) one dataset with multiple sensor measurements including EGG time series in 36 participants [25]. Also, three raw EGG data samples with software code for EGG signal analysis are shared by Wolpert et al. 2020 [19]. Hence, we would argue that an outstanding question of time series availability [2, 3, 26] is prioritized for EGG signals. To overcome the limited number of available EGG datasets, we propose a data augmentation approach to generate synthetic electrogastrogram time series (syEGG). Captivating mathematical model for investigating sources of normal and abnormal EGG data activities has been already developed [27, 28]. However, the method proposed in [27, 28] uses a torso-trunk model focused on signal distribution in relation to the stomach anatomy aiming to assess gastric slow wave propagation with multi-channel recording. Rather than studying origin of EGG signal and its distribution in trunk, we aim at producing syEGG. Unlike GAN, our augmentation method does not require large input data for the training phase and enables direct control of parameters related to EGG data. The model is motivated by real signals and a previously proposed pioneering simple dynamic model for generating synthetic ECG signals [29]. Presented augmentation method allows tuning of important parameters related to EGG data and shape of power spectral density (PSD) that faithfully present EGG signal characteristics further inspired by the approach applied in [15, 30].

The main objective in this study is to present a new data augmentation approach for generating syEGG. Available authentic data and known EGG data properties resembling normal gastric rhythm in healthy adults during postprandial and fasting states, as well as signal-related changes during simulator sickness occurrence are used to synthesize syEGG. Realism of produced syEGG-based features is validated statistically and in comparison, with the real-life parameters derived from EGG signals. The AI (artificial intelligence)-based approach for future application and validation of the presented method is proposed. The software code is freely shared via GitHub and Zenodo under the GNU General Public License (GPL) license to encourage further adoption and adaptation by other scientists and algorithm developers [31]. To the best of our knowledge, this is the first attempt to produce syEGG by a data augmentation method. Some aspects during the development of the proposed method are tested in the master thesis of Milenić [32].

## 2 Methods

All processing steps are performed in GNU Octave (GNU Octave, version 6.4.0. Copyright (C), The Octave Project Developers) [33]. We use the following GNU Octave packages: signal [34], statistics [35], communications [36], and fuzzy-logic-toolkit [37].

### 2.1 Simulation of normal syEGG rhythms in healthy adults

Electrogastrography procedure is used to capture electrical activity of the stomach smooth muscles. The resulting EGG

---

[1] PhysioNet: The Research Resource for Complex Physiological Signals, https://physionet.org/, Accessed on November 24, 2023.

[2] EGG abbreviation is used for both electrogastrography (a method to record electrical activity of stomach smooth muscles) and electrogastrogram (the resulting signal during electrogastrography procedure) in literature [19, 20]. To avoid ambiguity and to emphasize that we refer to electrogastrogram we use "EGG data", "EGG signal", and "EGG time series" throughout the paper.

[3] Databases presented in this paper are found through DatasetSearch by Google (https://datasetsearch.research.google.com/, Accessed on November 24, 2023) with the keyword "electrogastrography".

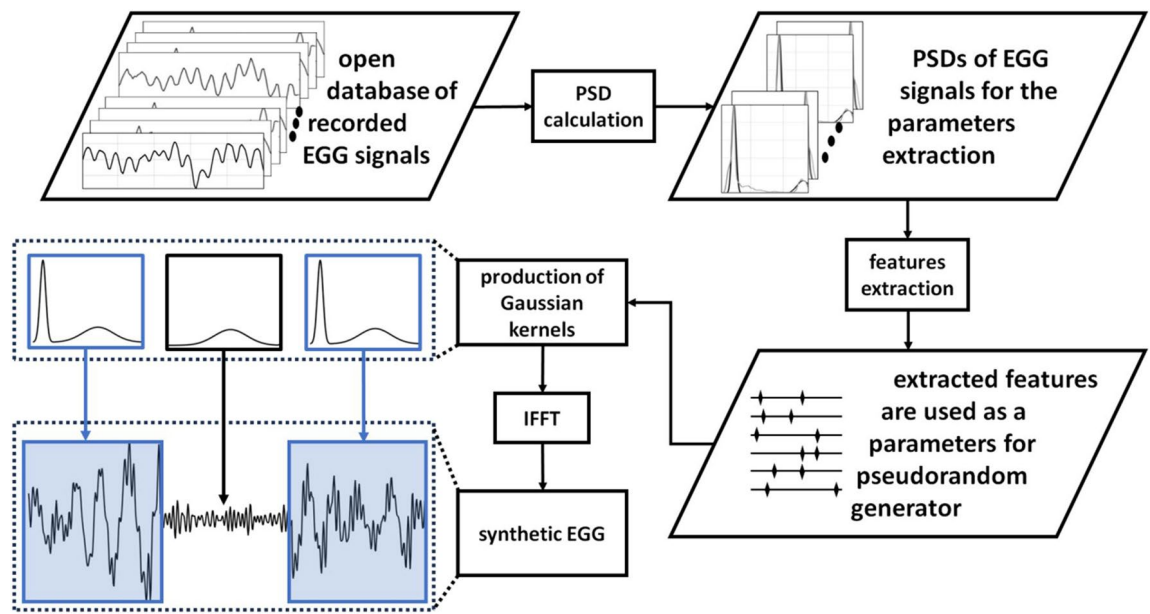

**Fig. 1** Block diagram of proposed methodology for generating simulated electrogastrograms (EGG signals) with features extracted from real-life signals from an open database. Abbreviations: PSD: power spectral density; IFFT: inverse fast Fourier transform

signal is measured as the difference of electrical potentials acquired between two recording surface electrodes. Essentially, EGG data characteristics reveal the stomach rhythm displaying normal (normogastria), fast (tachygastria), low (bradygastria), or absent (arrhythmia) stomach activity. Any rhythm in EGG signal that deviates from the normal rhythm is called dysrhythmia. Rhythms in EGG signals may be classified as deterministic stationary temporal series for short time intervals as EGG signal displays basic gastric rhythms in such circumstances which is a main hypothesis for the proposed methodology. Arrhythmia could be represented as a random stationary segment corresponding to the spontaneous and unspecific electrical activity in living tissue. Yet, EGG signal in healthy adults recorded during longer time intervals is non-stationary as it consists of normal rhythm, dysrhythmia, and arrhythmia. Therefore, EGG signal recorded during longer intervals is an expectedly random, complex, and multi-component signal revealing more than normal gastric rhythm that is commonly present with ≥ 70% of overall recording time in healthy adults. Main obstacle for wider EGG data adoption is its vulnerability to noises and particularly to movement artifacts, so the generation of clean synthetic EGG signals without artifacts could be a viable approach in evaluation of processing workflows for noise elimination [15, 20, 38, 39].

To properly mimic characteristics of real-life EGG signals, we modeled syEGG by power spectral density (PSD) features extracted from the available EGG signals recorded in fasting and postprandial states in healthy subjects. Then, simulated EGG data, i.e., dominant frequencies (DFs) are generated by the Monte Carlo simulation to test the existence of the statistical significance difference between DFs for fasting and postprandial states. Morphology of syEGG is captured from EGG signals available in an open access database and recorded during fasting and postprandial states from another publication [21, 22]. Detailed methodology is presented in Fig. 1.

The database consists of signals recorded in 20 healthy subjects by three-channel device with surface Ag/AgCl electrodes in two states (fasting and postprandial) resulting in overall 120 (20×3×2) time series segments with sampling frequency of 2 Hz (gain was set to 1000 and analog-to-digital converter resolution was 16 bits). Subjects were recorded for 20 min in the fasting state (after 6-h long fasting period with 2-h with no fluid intake prior to the measurement). Then, the test meal of about 300 kcal was provided to participants and EGG signals during post-prandial state (after meal intake) were recorded for another 20 min. During both recording sessions subjects were placed in supine position. Electrodes were placed between greater and lesser curvatures of the stomach with the aim to cover electrically active stomach area [21, 22].

Proposed data-driven approach is inspired by methods applied in [29, 40] and produces artificial PSD of EGG data by introducing Gaussian kernels. The simplest syEGG model realization resembles typical recording in a healthy person and incorporates two kernels—one corresponding to the normogastria and the other mimicking breathing artifacts. Namely, syEGG PSD $S(f)$ is generated with a sum of Gaussian functions. Inverse fast Fourier transform (IFFT) of

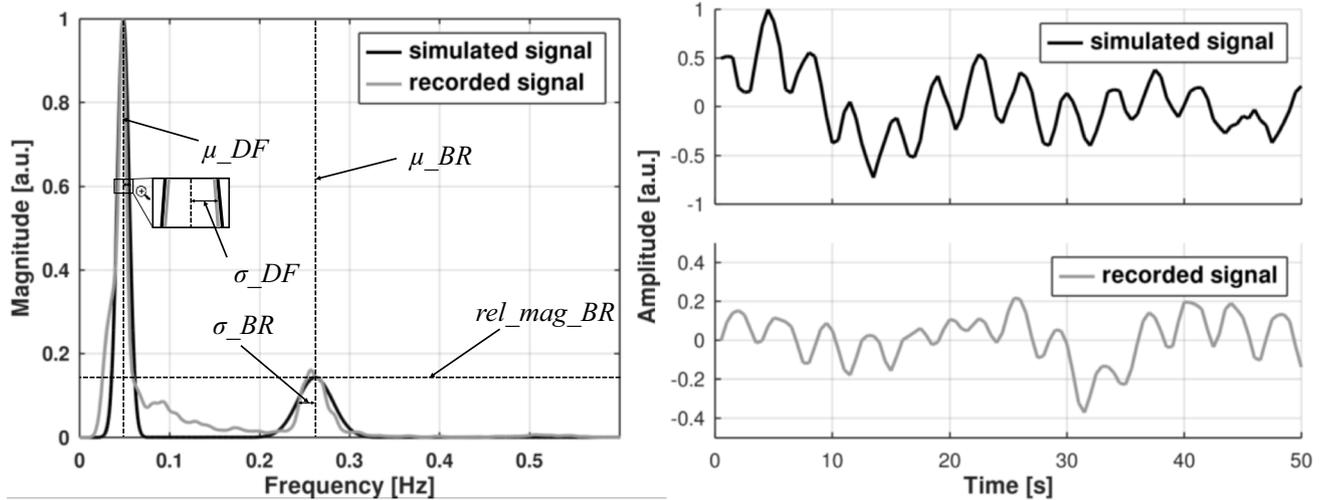

**Fig. 2** Power spectral density (PSD) of recorded EGG signal from the open-source database [10, 11] (ID18_postprandial) with estimated Gaussian kernels is presented in the left-hand panel with following parameters: $\mu\_DF$: mean dominant frequency; $\sigma\_DF$: mean standard deviation of Gaussian kernels; $\mu\_BR$: mean peak value from fitted breathing artifact kernels; $\sigma\_DF$: mean standard deviation of Gaussian breathing artifact kernels. The syEGG PSD incorporates filtered additive Gaussian noise (please, see text for more details). On the right-hand panel the corresponding generated synthetic EGG and real-life EGG signals in time domain are shown

sqrt($S(f)$) with phases pseudorandomly distributed between 0 and $2\pi$ is used to generate syEGG (white noise). PSD of syEGG is built with two Gaussian kernels with parameters obtained from the real-life PSDs of EGG signals in healthy sample: normogastria kernel is defined by mean $\mu\_DF$ and standard deviation (SD) $\sigma\_DF$, while breathing artifact kernel is determined by $\mu\_BR$ and $\sigma\_DF$ (Fig. 2, left-hand panel). Separate normogastria kernels are used for simulating syEGG recorded during fasting and during postprandial states. To introduce natural EGG data variability, standard deviations for all determined parameters are subsequently incorporated with a pseudorandom generator into the model. Also, a variation of the PSD model is introduced by a pseudorandom generator to mimic the real-life PSD of EGG signals. The scale of PSD variation resembling overall noise contamination can be increased by the user's choice; Fig. 2 presents a simulated signal without PSD variability (the scale is by default set to zero). Corresponding presentation of simulated and recorded signals from left-hand panel in Fig. 2 in time domain are shown on the right-hand panel in Fig. 2.

To model Gaussian kernel resembling normogastria, we firstly filter available EGG data with the 3rd-order Butterworth band-pass filter with cut-off frequencies of 0.03 Hz and 0.6 Hz to remove noises and artifacts outside of the spectral range of normogastria [21], but wide enough to include breathing artifacts. The filtering is performed in both directions to achieve zero-phase filtering which results in doubling the filter order (final filter order is 6th order). We select Butterworth filter for its flat magnitude response [41]. Although Butterworth filter can produce ringing artifacts in step response, the use of zero-phase filtering reduced these artifacts.[4]

Then, we estimate Welch's PSDs of 120 filtered EGG time series (window size is set to 300 samples, i.e., 12.5% of the signal length, with overlap of 50%). Then, DFs are determined for each EGG time series and the mean DF is used to estimate separately $\mu\_DF\_Fast$ and $\mu\_DF\_Post$ for fasting and postprandial states, respectively. All PSDs of EGG data are fitted with Gaussian curves in normogastria range 2–4 cycles per minute (cpm) [42]. to obtain fitted SDs for each time series and the mean SD for fasting ($\sigma\_DF\_Fast$) and postprandial states ($\sigma\_DF\_Post$). A Gaussian curve mean is fitted as a mean of the PSD while its SD is fitted as PSD standard deviation, weighted by PSD magnitude.

Then, representative 10 PSDs of EGG data (fasting IDs 2, 17, and 19; postprandial IDs 1, 3, 4, 9, 17, 18, and 19) are manually selected. The criterion for selection of distinguished breathing artifact contamination in spectral domain is judged by the visual observation to ensure that only time series with expressed and clear breathing artifacts are used. To estimate kernel, mean $\mu\_BR$ we averaged PSD weighted mean values in narrow breathing artifact range (0.2 to 0.4 Hz). Next, to estimate kernel SD $\sigma\_BR$, we average SDs from individual Gaussian curve fits. To the best of our knowledge, there is no documented difference between

---

[4] Following kind suggestion from anonymous Reviewer, we performed detailed qualitative and quantitative evaluation of ringing artifacts in Butterworth filter that justify our filter selection. The code for additional analysis is available at GitHub in GNU Octave file stepAnalysis.m [31].

**Table 1** Simulation parameters for PSD Gaussian kernels of syEGG in fasting and postprandial states, as well for Gaussian kernel of breathing artifact. Natural variability of obtained parameters is simulated with a pseudorandom number generator with Gaussian distribution (mean and SD columns)

| Parameter | Kernel | Explanation | Mean | SD |
| --- | --- | --- | --- | --- |
| $\mu\_DF\_Fast$ | Normogastria | DF in fasting state | 2.9336 cpm | 0.1094 cpm |
| $\mu\_DF\_Post$ | | DF in postprandial state | 2.9743 cpm | 0.1158 cpm |
| $\sigma\_DF\_Fast$ | | SD of Gaussian kernel in fasting state | 0.4836 cpm | 0.0740 cpm |
| $\sigma\_DF\_Post$ | | SD of Gaussian kernel in postprandial state | 0.4794 cpm | 0.0823 cpm |
| $\mu\_BR$ | Breathing artifact | Breathing peak | 16.7410 cpm | 1.0628 cpm |
| $\sigma\_BR$ | | SD of Gaussian kernel of breathing artifact | 2.6655 cpm | 0.4919 cpm |
| $rel\_mag\_BR$ | | Relative magnitude of breathing Gaussian kernel | 0.1907 a.u | 0.2474 a.u |

Abbreviations: *DF* dominant frequency, *SD* standard deviation, *a.u.* arbitrary unit, *cpm* cycles per minute breathing patterns before and after meal intake and respiratory artifact kernel is used for simulating syEGG during both fasting and postprandial states. Overview of means and SDs of determined kernels parameters is presented in Table 1. Generated PSD is normalized so that the peak of normogastria has a magnitude of 1, while PSD of the breathing artifact is scaled according to parameters in Table 1. Additive and positive colored noise is introduced to the artificial PSD to mimic its natural variability before applying square root and IFFT to generate artificial syEGG. To construct colored noise, firstly, magnitude samples are randomized between 0 and a user-specified input which is expressed as a fraction of the peak magnitude. Then, samples are filtered using a median filter with a window width of 1% of the full spectrum and the resulting-colored noise is added to the spectrum. Sample syEGG presented with real-life EGG signal for comparison during postprandial states are presented in Fig. 2.

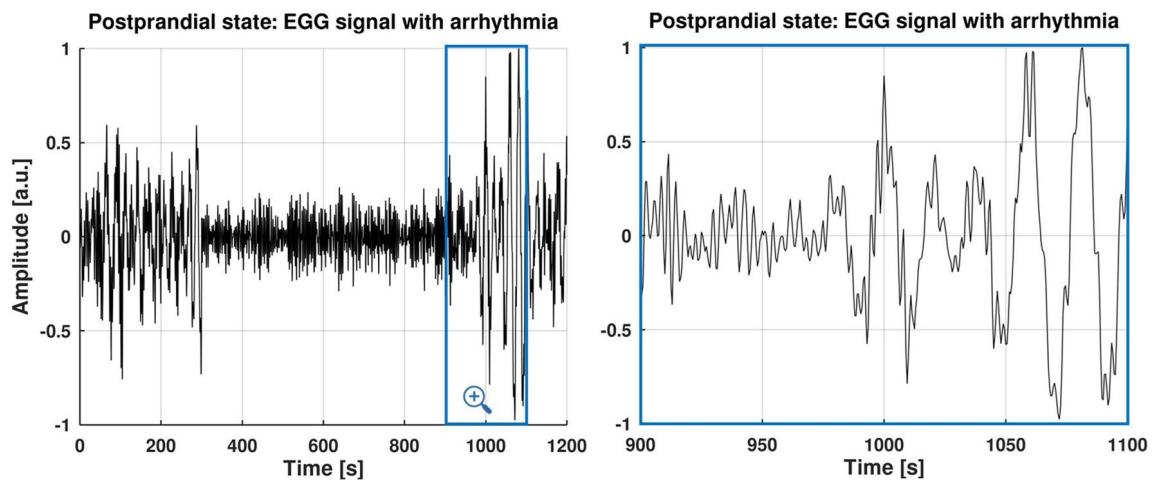

**Fig. 3** Time-domain representation of simulated EGG time series (syEGG) with an arrhythmia (absence of gastric rhythm) from 300 and 900 s

The absence of rhythm in simulated signal (syEGG arrhythmia) is modeled by combining a breathing artifact Gaussian kernel transformed in the time domain with a pseudorandom phase (white noise) and an additive-colored noise (please, see the time series snippet from 300 and 900 s in Fig. 3). In Fig. 3, EGG signal during arrhythmia has PSD that is flat for all frequencies as expected [43] as arrhythmia is mostly constructed of white noise, while single dominant peak resembles breathing artifact. To illustrate this, in software repository [31], we share a script (paper_figs.m) that plots two running spectrum analyses (RSAs) for arrhythmia: (1) with Welch's method with window width of 150 s and (2) with Yule-Walker's method (window width of 60 s). This segment is concatenated with rhythmical syEGG in the time domain resulting in heteroscedastic time series that possess unequal variability to properly mimic arrhythmia.

The syEGG model can be found in the syEGG.m GNU Octave function [31] where user can set the following parameters: (1) duration of the generated sequence (with default set to 1200 s), (2) sampling frequency in Hz (default value is 2 Hz), (3) recording state (postprandial or fasting state) with fasting being default, (3) the presence of breathing artifact contamination with default value of 1 indicating that the artifact is present (if user enters 0, the generated syEGG will not be added to the synthetic PSD), (4) a flag whether the output would produce plots, (5) seed for the sake of reproducibility (if not set to an integer, the value is selected pseudorandomly to warrant syEGG variability with each run), (6) arrhythmia occurrence with an array comprising start and end points in s, respectively for arrhythmia beginning and end (default points are 0 s), and (7) overall noise contamination with additive colored noise to produce proper PSD variability of EGG signal (the default scale is set to zero and can be further increased by user to the desired scale). The outputs of this function are as follows: (1) an array comprising time series of generated EGG signal, (2) PSD of generated signal, (3) DF in Hz, (4) the frequency in Hz of maximum of breathing artifact, (5) width of the DF peak, and (6) width of the breathing artifact peak.

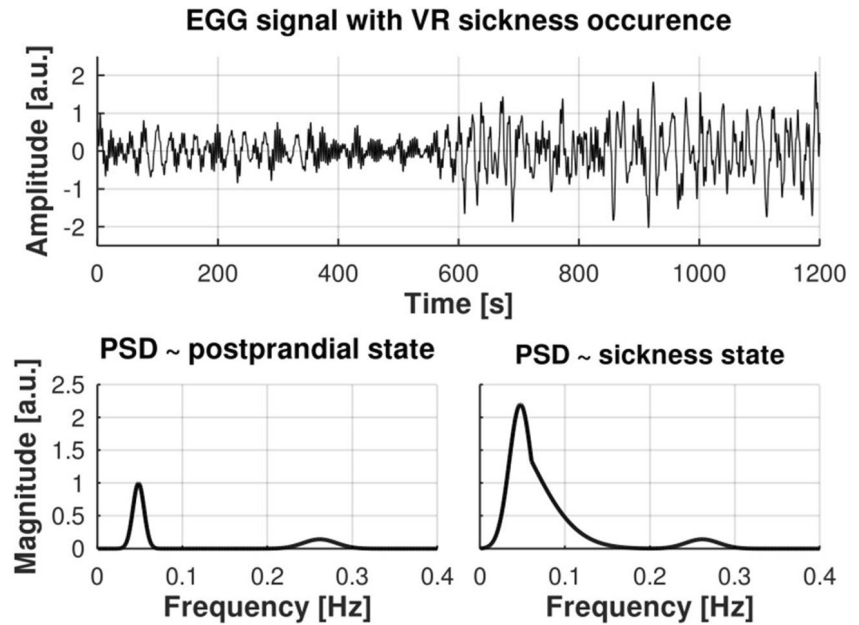

**Fig. 4** Synthetic EGG signal (syEGG) before and during simulator/VR sickness phenomenon with the corresponding power spectral densities (PSDs) during postprandial and sickness states. VR stands for virtual reality

### 2.2 Simulation of syEGG rhythms in healthy adults resembling sickness occurrence

Alteration from the normal electrogastrogram rhythm besides rhythm disturbances may include power variations as well. The corresponding abnormal gastric activity may indicate critical conditions such as functional dyspepsia, delayed gastric emptying, and idiopathic gastroparesis [39, 44]. Among other interesting applications, assessment of simulator sickness phenomenon caused by the subject's exposure to virtual reality systems or driving simulations can be performed with means of electrogastrography procedure, as EGG signal provides insight into gastric myoelectrical disturbances caused by nausea [39, 45, 46]. Therefore, we customize the proposed syEGG model to generate syEGG as a result of nausea induced by virtual reality experience and validate it against previously reported EGG time series properties.

For modeling syEGG during the simulator sickness occurrence and to mimic presence of dysrhythmia, we approximate the PSD with a combination of two Gaussian curves (Fig. 4). The phase spectrum is pseudorandomized and the waveform is obtained similarly to syEGG via IFFT. Such signals are then simply inserted into the waveform at specified positions following heteroscedastic models. The positions determine the time onset for sickness occurrence. We use visual presentation of PSD of EGG signal during nausea occurrence [45] for generating a model of syEGG during simulator sickness with the deductively derived parameters. During simulator sickness, magnitude of PSD is increased by a factor of 2.2. There is no formal basis for the magnitude increase (factor 2.2)—we used it provisionally having in mind that such changes can be from slight to drastic: in [45] ratios of total power (Table 2 in [45]) ranged from 1.83 to 3.52, while in [46] increase in RMS relative to the baseline was up to 50. Overall, it varies significantly among individuals. For synthetic signals, users can set any value or even further customize it. Dysrhythmia is modeled by stretching the normogastria Gaussian curve after the point of $\mu\_DF + \sigma\_DF$ (single standard deviation after the peak) horizontally by a factor of 4. Most commonly observed shift in PSD of EGG signal during simulator sickness [45, 46] is frequency increase into tachygastria range (up to 10 cycles per minute (cpm)), while normogastria is in range from 2 to 4 cpm. This fact is used to set a stretching factor at 4, but any factor from 2 to 5 would yield to a simulation of real-life event depending on the severity, i.e., prevalence of tachygastric rhythm in EGG signal. All models incorporate seeds for the sake of reproducibility.

**Table 2** Evaluation parameters for total power of PSD, percentages of power in PSD in three gastric rhythms (normogastria, tachygastria, and bradygastria), MF, and CS of PSD. All parameters are presented for two states (fasting and sickness states) in simulated EGG signal sequence of 1200-s duration along with expected changes reported in the literature for real-life EGG data. The simulation is repeated 100 times and mean values along with SDs are reported

| syEGG parameters | Fasting state | Sickness state | Expected change during sickness occurrence [43–46] |
| --- | --- | --- | --- |
| Total power of PSD (a.u.) | 0.02 ± 0.00 | 0.40 ± 0.23 | Increase |
| Normogastria (%) | 78.53 ± 7.01 | 33.38 ± 4.64 | Decrease |
| Tachygastria (%) | 6.06 ± 3.46 | 45.27 ± 3.55 | Increase |
| Bradygastria (%) | 15.40 ± 6.59 | 21.35 ± 3.31 | Increase or no changes |
| MF of PSD (Hz) | 0.05 ± 0.00 | 0.06 ± 0.00 | Increase |
| CF of PSD | 8.41 ± 0.72 | 5.35 ± 0.46 | Decrease |
| RMS (a.u.) | 0.31 ± 0.04 | 2.12 ± 1.30 | Increase |
| FSD (%) | 2.71 ± 0.44 | 8.67 ± 1.40 | Increase |

Abbreviations: *CF* crest factor, *MF* median frequency, *PSD* power spectral density, *syEGG* simulated electrogastrogram time series, *a.u.* arbitrary units, *RMS* root mean square, *FSD* percentage of PSD that has higher value of one quarter of maximum PSD magnitude, *SD* standard deviation

A separate GNU Octave function termed syEGG_VR.m is shared [31] for the generation of syEGG during sickness occurrence contains two additional parameters in comparison to the syEGG.m: (1) sickness onset (default value is 600 s) and (2) sickness offset (default value is 1200 s). The resulting time series of syEGG for default parameters of the syEGG_VR.m function except for the postprandial state and with seed set at 5 are presented in Fig. 4. Also, Fig. 4 comprises resulting PSDs of EGG signal before and during simulator sickness.

### 2.3 syEGG simulator with normal rhythm validation

We validate the proposed approach for syEGG generation by using paired sample *t*-tests as in [21] to compare generated DFs of syEGG in fasting state against DFs produced in postprandial state in Monte Carlo simulation by replicating the experiment for comparison of 20 DFs obtained during fasting and postprandial states a million times. The same simulation is repeated with 100 DFs. Furthermore, we present the fraction of the resulting *p* values smaller than three thresholds 0.05, 0.01, and 0.001 that showed statistically significant differences between DFs in syEGG for two states. The basis for Monte Carlo evaluation comes from the previously published analysis on used real-life data [21, 22] as we aimed to replicate analysis on simulated data to compare obtained results and appropriately assess syEGG.

### 2.4 syEGG simulator with sickness occurrence validation

To evaluate the realism of generated syEGG during VR-related sickness occurrence, we calculate following parameters proposed in [45]: (1) total power of PSD, (2) power in percentage for three EGG time series rhythmic ranges (normogastria, bradygastria, and tachygastria), (3) median frequency of PSD (MF), and (4) crest factor of PSD (CF) for the first half of the generated signal without sickness occurrence and for the second half during sickness occurrence. Both segments have the same duration of 600 s. Although, EGG data analysis is commonly performed in frequency domain [21, 45], some techniques are introduced with the aim of EGG signal analysis in time domain. Therefore, we add RMS of amplitudes during fasting and during VR-related sickness to assess syEGG in time domain [46]. Also, we calculate percentage of PSD that has higher value of one quarter of maximum PSD magnitude (FSD) for the completeness of presented features [46]. All calculated parameters are obtained from 100 repetitions of syEGG generation for sickness occurrence. The seed is changed through the for loop and for each iteration set to an iterator for obtaining reproducible results.

## 3 Results

The results of Monte Carlo simulation revealed that the percentages of *p* values of the paired sample *t*-tests for pairs of 20 DFs have 62.5%, 37.7%, and 12.4% realizations with *p* values smaller than 0.05, 0.01, and 0.001 thresholds, respectively. This means that in 62.5% cases of overall million simulations (625,000 cases) of 20 different DFs for postprandial state and 20 different DFs for fasting state, statistical significance at level 0.05 is reached. In other 375 000 cases, statistical significance was absent for comparing DFs in fasting and postprandial states. Similarly, when algorithm is repeated for 100 DFs a million times, proportion of 71.7%, 47.6%, and 21.1% realizations with *p* values is smaller than 0.05, 0.01, and 0.001 thresholds, respectively.

In Table 2, comparisons of syEGG parameters for two states with and without sickness (sickness and fasting states) occurrence are presented with mean values and SDs for all parameters. Also, Table 2 comprises expected change in relation to data from literature during the sickness occurrence for immediate comparison.

## 4 Discussion

It is worthwhile to mention that the proportion of 100% of *p* values being smaller than 0.05 would be too enthusiastic to anticipate having in mind that both DFs in simulated EGG signals during fasting and postprandial resemble normal gastric rhythm and have overlapping ranges (Table 1). Though 62.5% is mildly convincing, it is important to note that presented data augmentation approach does not take into account individual differences as we could not incorporate them in our model. In other words, we do not simulate separate subjects and their changes, as we rather generalize signal properties (DF parameter) for healthy individuals before and after meal intake. Also, we followed the number of simulated DFs of 20 as reported in [21] which is relatively low. For 100 DFs, the proportion of *p* values smaller than 0.05 increased to 71.7% expectedly. Utilized open dataset comprises EGG signals obtained from healthy individuals with both low and high body mass indexes (BMIs) of BMI < 25 kg/m$^2$ and BMI > 25 kg/m$^2$, respectively [21]. As we do not control for BMI in the proposed method for generating simulated EGG data, it could be expected that in cases of higher BMI, a statistically significant difference between fasting and postprandial may be absent as suggested in [21]. Additionally, we do not control for the recording site as electrode locations can influence statistically significant separation between fasting and postprandial states [21]. Therefore, 62.5% and 71.7% may be considered as good enough results, having in mind that we do not control for individual differences including BMI, as well as for electrode locations. Future studies may incorporate such effects depending on their final application.

We present a modification of the EGG signal generator to produce syEGG alterations that correspond to simulator sickness phenomenon. Though derived features showed expected trend (Table 2) [46], the PSD is modeled on the qualitative basis, unlike fasting and postprandial syEGG generators that are modeled quantitatively. Namely, we use visual representation of PSD from [45] to shape syEGG during simulator sickness occurrence. Moreover, we demonstrate that both frequency- and time-domain features revealed expected trends (Table 2). Specifically, it is interesting to note that RMS increased at large with much higher SD which perfectly corresponds to sickness-related EGG signal changes [45, 46]. The drawback of simulated signal with sickness occurrence is that we cannot calculate the sickness onset as we lack additional recording in the simulator that does not produce sickness sensation to appropriately replicate the method for onset computation proposed in [46]. Forthcoming proposed method application may include syEGG generation inside VR for different levels of sickness as well.

Moreover, future access to a database comprising recorded alterations in EGG data caused by the simulator sickness would enable calculation of model parameters empirically. These alterations would include also post simulator-related changes in EGG signal and consequently would comprise of more than two heteroscedastic segments. On the other hand, although fasting and postprandial states were simulated empirically, we could not simulate dysrhythmic gastric activity as available database in healthy subject resembled dominant normogastria. Further expansion of presented models with patients' data and pathological dysrhythmias would definitely improve the proposed data-driven model by customizing it for these specific cases. We would argue that successful customization for sickness occurrence reveals a firm basis for further tailoring of the presented model for different gastro-intestinal pathologies in EGG signal or even for producing other biosignals (e.g., electroencephalogram and electrohysterogram).

We followed the ideas pioneered by McSharry et al. [29] that were previously criticized for the lack of diversity and realism in the generated signal [26]. Despite the lack of dysrhythmias in the basic model and in the rather theoretical approach in modeling sickness phenomenon, we would argue that additional pseudorandomness of the parameters controlled by the seed in the proposed method provide more divergence for syEGG in comparison to the dynamic ECG model.

Since the main novelty in our algorithm is the simulation of EGG signals with control of signal parameters and recording states, we should start by taking a close look at EGG signal characteristics and nature. EGG signal has abrupt changes in amplitude and morphology as a result of simulator sickness [45, 46]. We introduce these points of change as defined in [30] for the EMG signal as the boundary points for data distribution alteration (heteroscedasticity). Such models introduced novel methods for calculation of EMG envelopes based on the activation and deactivation patterns constructed from these points of change [47]. The presented syEGG model could be further explored in a similar way to design proper tools for the reliable onset and offset detection of rhythm changes in EGG signal.

Although statistical methods are cited more in the literature for the EGG signal analysis, there is a growing trend of reassuring results from the field of artificial intelligence (AI) [38]. In general, large open access high-quality datasets are required to improve machine learning solutions in medicine, especially for better generalization [3, 5, 26]. Available medical datasets may foster the development of automated medical diagnostics and serve for educational purposes [26]. The presented approach may assist the growing area of machine learning in interpretation of EGG data, especially as data annotation can be done at the generation phase. Data augmentation is commonly applied to features [3]. Here, we aim at presenting syEGG mainly for the purpose of development of processing methods and understanding the underlying mechanisms, but final application of the produced artificial data is not limited only to the signal processing algorithms, as they can readily be applied in evidence-based medicine incorporating AI. Available human, realistic, and synthetic EGG signals would enable assessment of signal processing techniques in relation to the sampling frequencies, signal duration, noise levels, and morphological changes in EGG signal and syEGG would also facilitate training of machine learning algorithms, as well as comparisons among different algorithms and processing workflows ultimately leading to convincing scientific conclusions. Moreover, development of efficient algorithms would foster EGG signal processing standardization and its medical application. Modeling biosignals in general can also promote deeper and thorough understanding of the biological system and can provide an opportunity to study diverse health conditions [15].

To summarize, proposed method for generating syEGG is based on two known approaches: (1) PSD-based method to model the PSD of stationary syEGG (with respiration artifact) inspired by McSharry et al. [29] as they used the same procedure for modeling RR-interval process and (2) heteroscedastic model to introduce non-stationarity in EGG time series resembling rhythmical and arhythmical stomach activity adopted by Guerrero and Macías-Díaz to represent electromyogram activation/deactivation patterns in time domain [30, 44]. We would argue that the main novelty of proposed approach lies in the fact that the shape of PSD can be altered so that it matches changes in EGG signal during pathological states such as VR-induced sickness (as we investigated earlier in [45]) by using a combination of two Gaussian shapes.

### 4.1 Limitations of the study

A new method for producing syEGG is introduced. We proved its efficacy in producing realistic synthetic EGG signals. However, the proposed solution can be further advanced by the following:

1. Introduction of additional noise generators. Specifically, we incorporate only normally distributed noise and depending on the sampling frequency syEGG can be contaminated with other noises as for example with heart pulses or with dynamic and static noises.
2. We used statistical methods to evaluate synthetic EGG time series, but both statistical and ML-based methods may be used to provide appropriate differentiation between real and synthetic data [6, 12].
3. Future upgrades can incorporate spike potentials as the syEGG method simulates only stomach electrical control activity (ECA) or the so-called slow waves, but does not incorporate electrical response activity (ERA), i.e., spike potentials [48].
4. IFFT could be replaced by the inverse wavelet transform with for instance Daubechies wavelet basis to explore whether it would provide more realistic and computationally efficient syEGG as it has been previously proposed for heart rate variability [40].
5. More Gaussian kernels could be used to mimic different rhythms in EGG signal (this is partially explored by qualitative method used for simulating EGG data during VR experience and sickness occurrence). Moreover, additional kernels would enable different approach in realization of percents of normogastria in the simulated time series.
6. White noise is commonly used to mimic natural variability of electrical signals in living tissues [49]. Here, we proposed an additive-colored noise due to its similarity to natural EGG data variability determined by visual inspection. The estimation of natural variability is out of scope of this paper, but future research may be directed towards the estimation of the exact type of noise and contamination level presented in the PSD of EGG signal resembling PSD variability.
7. Adaptation for real-time simulators of analogue signals based on microcontrollers could be the future direction as proposed in [4, 50, 51], but we did not consider those here. Computational load imposed on the processor should be considered in such cases. Also, computational load is important in when there is a need for generation of large amount of data (e.g., for ML algorithms), but this can be done in a separate session and data can be saved for offline use. To provide an insight into computational load, we assessed the processor total time spent executing the GNU Octave session. For execution of *syEGG()* with default parameters we obtained $0.80 \pm 0.05$ s (1000 repetitions) for processor time. Total time was slightly higher by ~4% when Mozilla explorer (8 tabs) and one LibreOffice document were active ($0.83 \pm 0.07$ s). When the total number of generated samples increased from 2400 to 4800, total processor time increased to $1.08 \pm 0.03$ s by 35%. For generating two times less: 1200 samples (600 s) total time is decreased to $0.66 \pm 0.02$ s by ~17%. There is still room for improvement as proper optimization to reduce computational complexity would result in faster execution, but this is out of scope of this paper.

8. We aim to allow control for common EGG data parameters (e.g., DFs), but future design could incorporate non-linear optimization techniques to fit the PSD into the model as proposed in [52].
9. To properly assess usability of the proposed model for generating synthetic data, it would be desirable to train an AI algorithm for sickness classification with and without synthetic EGG data in the training set and then to compare AI evaluation metrics. We could not perform such evaluation as we have only one real-life dataset at our disposal for healthy individuals consisting of EGG signals recorded in fasting and post-prandial states [21, 22]. Future work on evaluation of AI algorithms trained on the synthetic data and tested on real-life data would be highly welcomed for proper recognition of syEGG possibilities. Additionally, data manipulation by generation of synthetic datasets or other types of data alteration techniques could be employed for building Explainable Artificial Intelligence (XAI) models [53].
10. For assessing appropriate PSD parameters, we use all three channels from one recording session to include as much as possible recorded signals in our analysis. Although in electrophysiological sense information from multiple recording sites represent the same information, signals from multiple channels are not the same as conduction delay takes place as noted by other research groups using multi-channel recordings in electromyography studies [54, 55]. In EGG studies, there are two main approaches for recording multi-channel EGG signals [19]: (1) to select one recording site with the least noise contamination for further simple single-channel analysis and (2) to study the wave propagation as proposed in [56–58]. Here, we aimed at the first approach. In initial study that resulted in the used open dataset [21, 22], it is shown that recording sites differ in relation to BMIs of healthy subjects which has been revealed by cross-correlation coefficients. Future adaptations of the proposed method to generate syEGG on a larger dataset may yield to a better fit of simulated EGG data to real-life signals.
11. Phase of the syEGG is selected to be of random nature. We adopt rationale applied by McSharry et al. for generating synthetic heart rate variability [29] with the assumption that phase will not affect the realism of the signal as EGG signal does not contain fiducial points as for example ECG signal. Overall, biomedical signal processing methods rarely use phase information for feature extraction [59]. Intelligibility of phase-only reconstruction of images, crystallographic structures, and waveforms (speech) has been already demonstrated [60]. However, this is only possible for long sequences of speech and under certain assumptions as elimination of magnitude deteriorates many important aspects of signal [60, 61]. In our case, we hypothesize that phase is unimportant since we generate relatively short intervals of stochastic signal with assumption of stationarity and as phase plots of available EGG signals revealed that the phase appears to be of random nature. Non-stationary parts of the syEGG are added following applied heteroscedastic model in time domain. Finally, ss phase preserves the location of events in time (for one dimensional waveform such as EGG time series) and space (in case of images it is edge information useful for contouring) [60], we use random phase and add location of events, such as arrhythmia and sickness occurrence in time domain by adopting heteroscedastic model. Other modifications of the proposed method may yield different approach(es) in simulation of phase of EGG signal.
12. We use visual observation to select records of EGG signals with prominent breathing artifacts. Manual interventions like visual observation can jeopardize the computational reproducibility of EGG signal analysis workflows, but educated observation may be necessary for the analysis of artifacts in EGG signal [62–66]. In an attempt to develop automated procedure for selection of prominent breathing artifacts, we designed a simple procedure shared at [31] (script: automaticDetectionBA.m). Although method that we developed automatically outputs records of EGG signals in descending order by comparing PSD ranges of EGG signals and breathing artifacts, further refinement is needed to eliminate EGG signals with larger noise contamination. Therefore, we decided to keep educated visual observation rather than automated procedure.
13. Larger database with EGG signals recorded during longer periods of time and during different protocols is required for further confirmation or improvement of the proposed data-driven approach for generating syEGG.

## 5 Conclusion

We showcase a promising benchmark, synthetic, and natural-like synthetic EGG data representative that could be used for the development of novel and advanced machine learning algorithms, as well as for testing the robustness and other properties of existing processing techniques and methods for feature extraction and noise rejection. Besides generation of synthetic EGG signals with dominant normal rhythm during fasting and postprandial states which represent a typical finding in healthy adults, we present syEGG changes related to the simulator sickness occurrence as a demonstrative approach for further utilization of the presented method in case of normal gastric rhythm alterations and in relation to the gastro-intestinal pathologies. With the control of input parameters, the user can simulate other gastric abnormalities. The method for syEGG generation is available in open access. Availability of data-driven model to produce synthetic EGG data may tackle the growing problem of open biomedical data shortage.

**Acknowledgements** We kindly thank Prof. Sašo Tomažič from the Faculty of Electrical Engineering, University of Ljubljana for his valuable advises on signal processing methods.

**Author contribution** Nadica Miljković: conceptualization, formal analysis, and writing—original draft. Nikola Milenić: software, validation, and visualization. Nenad B. Popović: methodology, software, and visualization. Jaka Sodnik: methodology, investigation, and writing—review and editing.

**Funding** Nadica Miljković acknowledges the support from the grant No. 451–03-65/2024–03/200103 funded by the Ministry of Science, Technological Development and Innovation, Republic of Serbia. Jaka Sodnik acknowledges the support from Slovenian Research Agency under grant No. P2-0246. The Funders were not involved in the study design, collection, analysis, and interpretation of data; in the manuscript preparation; and in the decision to submit the manuscript.

**Code availability** The source code for generating synthetic EGG signals is openly available and shared with GNU GPL license ver. 3.0 at GitHub (https://github.com/NadicaSm/syEGG) [31].

## Declarations

**Ethics approval and consents** Not applicable.

**Competing interests** The authors declare no competing interests.

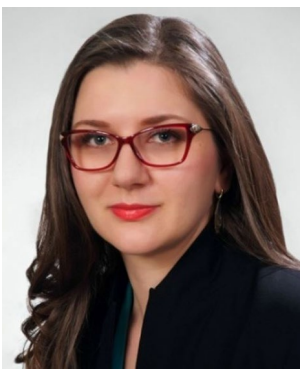

**Nadica Miljković** is Associate Professor of Biomedical Engineering at the University of Belgrade – School of Electrical Engineering and Guest Researcher at the Faculty of Electrical Engineering, University of Ljubljana. Her experience includes R&D of medical instrumentation (she was involved in developing novel medical rehabilitation devices) and her research is mainly focused on electrophysiological signal acquisition and biomedical signal processing both in patients and in healthy subjects .

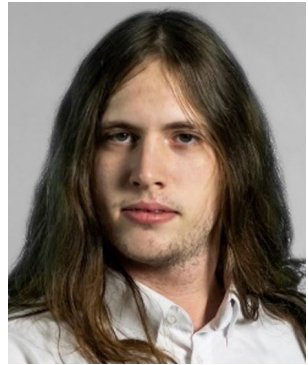

**Nikola Milenić** has received a Bachelor degree in electrical engineering and computer science at the University of Belgrade—School of Electrical Engineering in 2020, where he also pursued a Master degree in 2023. His research interests include biomedical signal processing and high performance computing (HPC). He also works as hardware engineer at NextSilicon, developing HPC processors.

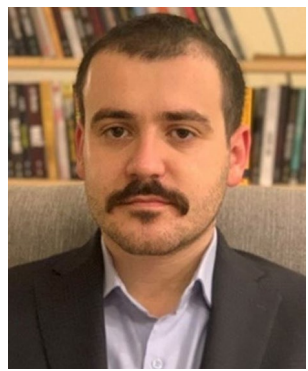

**Nenad B. Popović** finished his bachelor, master, and doctorate studies at the School of Electrical Engineering, University of Belgrade. The main areas of his research are the acquisition, processing, and application of electrophysiological signals. The focus of his research and the topic of his PhD Thesis was electrogastrography in which he explored recording procedures, signal processing, and feature extraction. His expertise includes engineering aspects of cardiac electrophysiology as he works as a technical consultant for pacemakers and implantable cardioverter defibrillators .

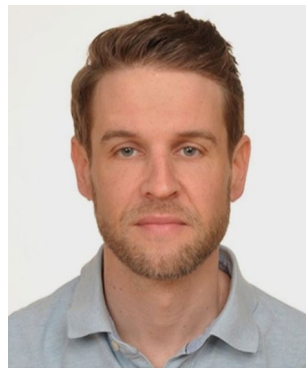

**Jaka Sodnik** is a Professor for the field of Electrical Engineering, at the Faculty of Electrical Engineering, University of Ljubljana. His research focuses on human factor and human–machine interaction in vehicles, driver behaviour and assessment of driving performance. He leads several national and international research projects and collaborative projects with industrial (Renault, Ford, Toyota, ISID, Virtual Vehicle, AMZS) and academic partners (University of Washington, Virginia Tech, Stanford, CARISMA, PLUS) focusing on methods for driver evaluation and profiling as well as mechanisms for assessing performance and safety of autonomous vehicles .